\documentclass[prc,superscriptaddress,showpacs,showkeys,twocolumn]{revtex4}%
\usepackage{graphicx, color, latexsym, amssymb, amsmath, multirow, mathrsfs, booktabs, CJK}
\usepackage[section]{placeins}%
\usepackage{amsfonts}%

\providecommand{\U}[1]{\protect\rule{.1in}{.1in}}

\newcommand{\beq}{\begin{equation}}
\newcommand{\eeq}{\end{equation}}
\newcommand{\beqn}{\begin{eqnarray}}
\newcommand{\eeqn}{\end{eqnarray}}
\newcommand{\bea}{\begin{array}}
\newcommand{\eea}{\end{array}}
\newcommand{\bsub}{\begin{subequations}}
\newcommand{\esub}{\end{subequations}}
\newcommand{\bpm}{\begin{pmatrix}}
\newcommand{\epm}{\end{pmatrix}}

\newcommand{\RE}{{\rm E}}
\newcommand{\RD}{{\rm D}}
\newcommand{\RV}{{\rm V}}
\newcommand{\RVT}{{\rm VT}}
\newcommand{\RT}{{\rm T}}
\newcommand{\RPV}{{\rm PV}}
\begin{document}

\title{Nuclear halo structure and {pseudo-spin} symmetry}
\author{Wen Hui Long}\email{longwh@lzu.edu.cn}
\affiliation{School of Nuclear Science and Technology, Lanzhou University, 730000 Lanzhou, China}
\affiliation{Physik-Department der Technischen Universit\"at M\"unchen, D-85748 Garching, Germany}
\affiliation{School of Physics, Peking University, 100871 Beijing, China}
\affiliation{Department of Physics, Texas A\&M University, Commerce, Texas 75429, USA}
\author{Peter Ring}
\affiliation{Physik-Department der Technischen Universit\"at M\"unchen, D-85748 Garching, Germany}
\author{Jie Meng }
\affiliation{School of Physics, Peking University, 100871 Beijing, China}
\author{Nguyen Van Giai}
\affiliation{CNRS-IN2P3, UMR 8608, F-91406 Orsay Cedex, France}
\affiliation{Univ Paris-Sud, F-91405 Orsay, France}
\author{Carlos A. Bertulani}
\affiliation{Department of Physics, Texas A\&M University, Commerce, Texas 75429, USA}

\begin{abstract}
Nuclear halo structure and {conservation} of relativistic symmetry are studied within the framework of the relativistic Hartree-Fock-Bogoliubov (RHFB) theory. Giant halos as well as ordinary ones are found in Cerium isotopes close to the neutron drip line. Bridged by $T=0$ {channel}, the {conservation} of pseudo-spin symmetry (PSS) plays an essential role in stabilizing the neutron halo structures. The Fock terms, especially the $\rho$-tensor couplings, not only play significant role in the PSS {conservation} but also present substantial contributions to the $T=0$ {channel}, from which is well demonstrated the necessity of Fock terms.
\end{abstract}

\maketitle

Since the neutron halo was first proven to exist in $^{11}$Li \cite{Tan:1985}, unexpected exotic modes have intensively challenged our understanding of exotic nuclei with extreme neutron-to-proton ratio, that play an essential role in the evolution of the universe. As the typical exotic mode, nuclear halo {(see Refs. \cite{Howard:2000, Mueller:2007, Rotival:2009a, Rotival:2009b} and references therein)} - the extremely diffuse matter - may strongly enhance the reaction cross section, which is of special significance for the element synthesis in astrophysics as well as in the discovery of new superheavies. Accompanying nuclear halo occurrence, shell quenching \cite{Dobaczewski:1994, Chen:1995} has been found when approaching these exotic regions, e.g., the $N=8$ shell in Li.

The nucleon-nucleon (NN) interaction is originally due to meson exchange processes as predicted by Yukawa \cite{Yukawa:1935}. Within such approach the nuclear binding is achieved mainly by the equilibrium between the scalar (phenomenological $\sigma$) and vector ($\omega$) meson fields \cite{Serot:1986} whereas the tensor forces due to $\rho$ and $\pi$ meson exchanges take part in the shell structure evolution \cite{Otsuka:2005}. As a relativistic symmetry in the Dirac equation \cite{Ginocchio:2005}, the pseudo-spin symmetry (PSS) \cite{ Arima:1969, Hecht:1969} is an important general feature in the nuclear energy spectra. Its origin is due to a Lorentz scalar potential ($\sigma$ field) and a Lorentz vector potential (time component of $\omega$ field) equal in magnitude, but opposite in sign \cite{Ginocchio:1997}. In fact, the competition of large scalar and vector fields in nuclei explains naturally the spin-orbit potential \cite{Meng:2006}, which makes the exploration in exotic regions more reliable.

In exotic nuclei the valence neutrons or protons are loosely bound and their coupling with the continuum becomes {important}. In terms of the Bogoliubov quasi-particles, the relativistic Hartree-Bogoliubov (RHB) theory \cite{Meng:1998a, Vretenar:2005, Meng:2006} provides a unified and self-consistent description of both mean field and pairing {correlations} and automatically takes continuum effects into account. Besides $^{11}$Li \cite{Meng:1996}, the halo {phenomena} have been predicted by the RHB theory in Ne \cite{Poschl:1997, Lalazissis:1998NPA}, Na \cite{Lalazissis:1998NPA, Meng:1998PLB}, Ca \cite{Meng:2002PRC} isotopes. Giant halo structures in Zr isotopes have also been studied within the RHB framework \cite{Meng:1998PRL} as well as in the non-relativistic Hartree-Fock-Bogoliubov method \cite{Grasso:2006PRC}.

While limited by the Hartree approach, important ingredients such as the \emph{spin-dependent} tensor forces are missing in the RHB theory. In the density dependent relativistic Hartree-Fock (DDRHF) theory \cite{Long:2006, Long:2007}, the tensor forces due to $\pi$ and $\rho$ meson exchanges can be naturally taken into account and have brought significant improvements on the consistent description of shell evolution \cite{Long:2008} and appropriate conservation of PSS \cite{Long:2006PS, Long:2007}. In this {work}, we use Cerium isotopes to study the nuclear halo phenomenon and relevant {conservation} of PSS within the relativistic Hartree-Fock-Bogoliubov (RHFB) theory \cite{Long:2009b}, an extension of DDRHF. For Cerium isotopes the proton number $Z=58$ is closely related to the pseudo-spin partner states $\pi1\tilde f$ ($\pi1g_{7/2}$ and $\pi2d_{5/2}$){, which represent well conserved pseudo-spin symmetry either from the experimental data \cite{Nagai:1981} or from the theoretical calculations \cite{Long:2009}. As we will see, the corresponding} PSS {conservation} is essential for the neutron shell effects when approaching the neutron drip line.

\begin{figure*}[ht]
\includegraphics[width = 0.95\textwidth]{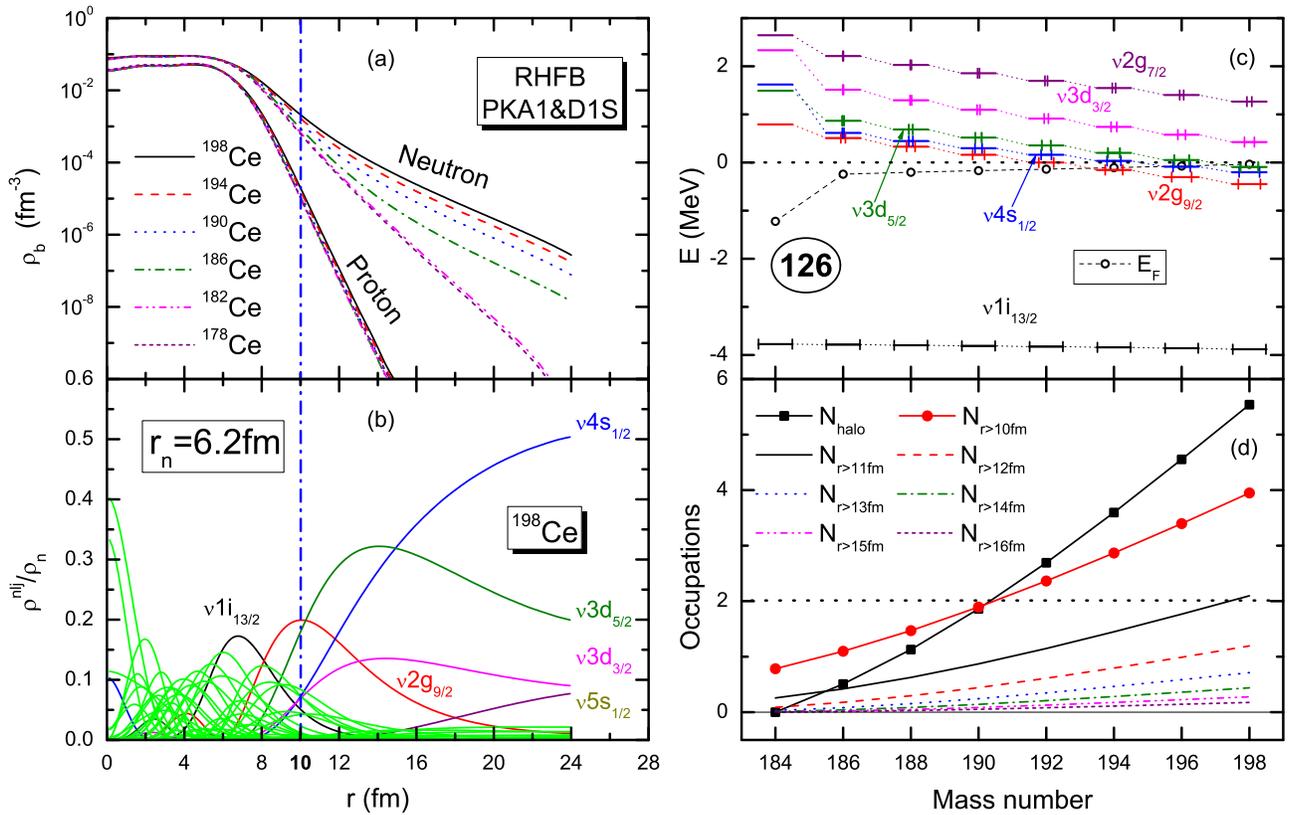}
\caption{(color online) (a) Neutron and proton densities, (b) the relative contributions of different orbits to the full neutron density in $^{198}$Ce, (c) neutron canonical single-particle energies {occupation probability (in $x$-error bars)} and Fermi energy $E_F$ {(in open {circles})}, and (d) {neutron numbers filling in the halo orbits $4s_{1/2}$, $3d_{5/2}$, and $3d_{3/2}$ ($N_{\text{halo}}$), and the ones lying beyond the spheres with the radii $r=10$, 11, 12, 13, 14, 15, 16fm, respectively $N_{r>10\text{fm}}$, $N_{r>11\text{fm}}$, $N_{r>12\text{fm}}$, $N_{r>13\text{fm}}$, $N_{r>14\text{fm}}$, $N_{r>15\text{fm}}$ and $N_{r>16\text{fm}}$}. The results are calculated by RHFB with PKA1 \cite{Long:2007} plus the Gogny pairing force D1S \cite{Berger84}. The spherical box radius is adopted as $R_{\text{max}} = 28$fm.} \label{fig:Density-SDG}
\end{figure*}

In Fig. \ref{fig:Density-SDG} we show nuclear matter distributions (left panels) and neutron canonical single-particle configurations (right panels) for Cerium isotopes close to the drip line. {The neutron drip line calculated here with PKA1 \cite{Long:2007} is $N=140$ whereas the calculations with PKO1 \cite{Long:2006} and DD-ME2 \cite{Lalazissis:2005} predict a shorter one as $N=126$.} As shown in Fig.\ref{fig:Density-SDG}a, the neutron densities become more and more diffuse after the isotopes $^{186}$Ce ($N=128$), a direct and distinct evidence of halo occurrence. From Fig. \ref{fig:Density-SDG}b we can see that such extremely extensive matter distribution, e.g. in $^{198}$Ce, is mainly due to the low-$l$ states, namely the halo orbits $\nu4s_{1/2}$, $\nu3d_{5/2}$ and $\nu3d_{3/2}$. Seen from the occupations of the halo orbits $N_{\rm halo}$ in Fig.\ref{fig:Density-SDG}d, the halos in $^{186}$Ce, $^{188}$Ce and $^{190}$Ce are ordinary, while $^{192}$Ce, $^{194}$Ce, $^{196}$Ce and $^{198}$Ce presumably have giant halos because more than two neutrons are occupying the halo orbits. {Similar conclusions can be also obtained from the neutron numbers lying beyond the sphere with the radius $r=10$fm ($N_{r>10\text{fm}}$ in Fig. \ref{fig:Density-SDG}d), which is large enough (the neutron matter radius $r_n=6.2$fm in $^{198}$Ce) for halos. Even extending to $r=16$fm, which is sometime taken as the radial cut-off in the calculations of stable nuclei, there are still some substantial amount of neutrons lying beyond this sphere in the isotopes from $^{192}$Ce to $^{198}$Ce.}

In Fig.\ref{fig:Density-SDG}c we find that the halo orbits ($\nu4s_{1/2}$, $\nu3d_{5/2}$ and $\nu3d_{3/2}$) are located around the particle continuum threshold where they are gradually occupied. For the isotopes beyond $^{184}$Ce ($N=126$), the Fermi levels (in open circles) approach the continuum threshold rather closely such that the stability of these halo isotopes becomes sensitive to pairing effects. Nearby the low-$l$ states, we find the high-$l$ states $\nu2g_{9/2}$ and $\nu2g_{7/2}$. Because of the relatively large centrifugal barrier for $g$-orbits they do not contribute much to the diffuse neutron distributions. Nonetheless, the existence of the high-$l$ states nearby halo orbits is still particulary significant, because it leads to a rather high level density around the Fermi surface, and evidently the pairing effects are enhanced to stabilize the halo isotopes.

\begin{figure}[htbp]
\includegraphics[width = 0.45\textwidth]{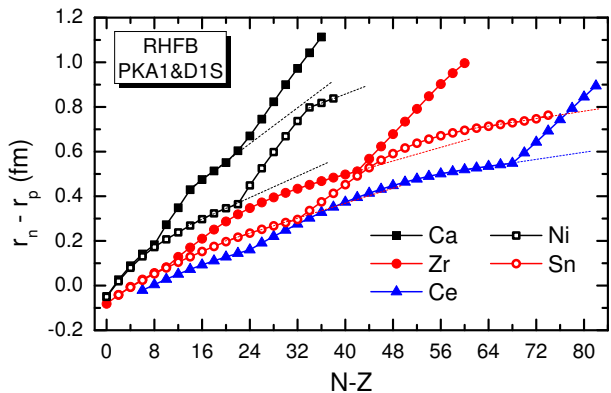}
\caption{(color online) Neutron skin thickness $r_{n}-r_{p}$ (fm) for Ca, Zr, Ni, Sn and Ce as a function of the isospin ($N-Z$). $r_n$ and $r_p$ are respectively the neutron and proton root mean square radii. The results are calculated by RHFB with PKA1 \cite{Long:2007}
plus the paring force D1S \cite{Berger84}. } \label{fig:PKA1}
\end{figure}

Evidence for the existence of a halo can also be found by studying the systematic behavior of nuclear bulk properties such as radii. In Fig. \ref{fig:PKA1} we show the isospin dependence of the neutron skin thickness ($r_{n}-r_{p}$) calculated in RHFB theory using the parameter set PKA1 for Ca, Ni, Zr, Sn and Ce isotopes. With respect to the behavior in the stable region (shown with the dashed lines), continuously growing deviations are found in the chains of Ca, Zr and Ce until the neutron drip line. This can be considered as evidence of a halo. Despite the deviations in the mid-region, Ni as well as Sn {show} an identical isospin dependence in both stable and neutron drip line regions, which may indicate only a neutron skin since the growth of a halo is interrupted. Compared with Ni, {Sn isotopes show} a much weaker skin effect.

\begin{figure}[htbp]
\includegraphics[width = 0.45\textwidth]{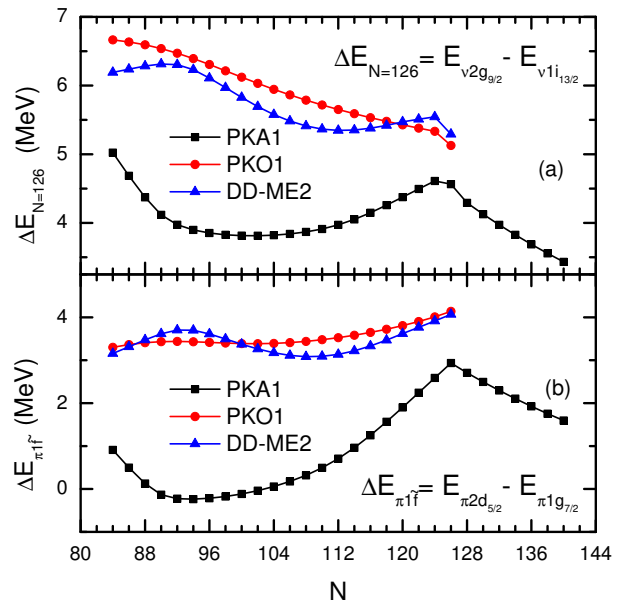}
\caption{(a) Neutron shell gap at $N=126$ ($\Delta E_{N=126} = E_{\nu2g_{9/2}} - E_{\nu1i_{13/2}}$) and (b) proton pseudo-spin orbital splitting ($\Delta E_{\pi1\tilde f} = E_{\pi2d_{5/2}} - E_{\pi1g_{7/2}}$) as functions of neutron number $N$ for Cerium isotopes. The results are calculated by RHFB with PKA1 \cite{Long:2007}, PKO1 \cite{Long:2006}, and by RHB with DD-ME2 \cite{Lalazissis:2005}. The Gogny force D1S \cite{Berger84} is adopted in the pairing channel.}
\label{fig:Ceshells}%
\end{figure}

Similar systematics as those of Fig. 2 are also found in RHB calculations for the chains of Ca and Zr \cite{Meng:1998PRL, Meng:2002PRC}. For the Ce isotopes the situation becomes quite different. RHFB calculations with the parameter set PKA1 (see Figs. \ref{fig:Density-SDG} and \ref{fig:PKA1}) show clear evidence for the existence of halo structures in the Ce isotopes, while in the calculations of RHFB with PKO1 \cite{Long:2006} and RHB with DD-ME2 \cite{Lalazissis:2005}, the isotopic chain ends at $N=126$, before the halo occurrence predicted by RHFB with PKA1. This deviation between models can be preliminarily interpreted by the shell structure evolution in Fig. \ref{fig:Ceshells}a, where much stronger shell effects are provided with PKO1 and DD-ME2 than with PKA1. As shown in Fig.\ref{fig:Density-SDG}c, {the neutron shell gap ($N=126$) between $\nu1i_{13/2}$ and $\nu2g_{9/2}$ states is close to the particle continuum threshold, which might essentially influence the stability of the drip line isotopes.}

\begin{figure}[htbp]
\includegraphics[width = 0.45\textwidth]{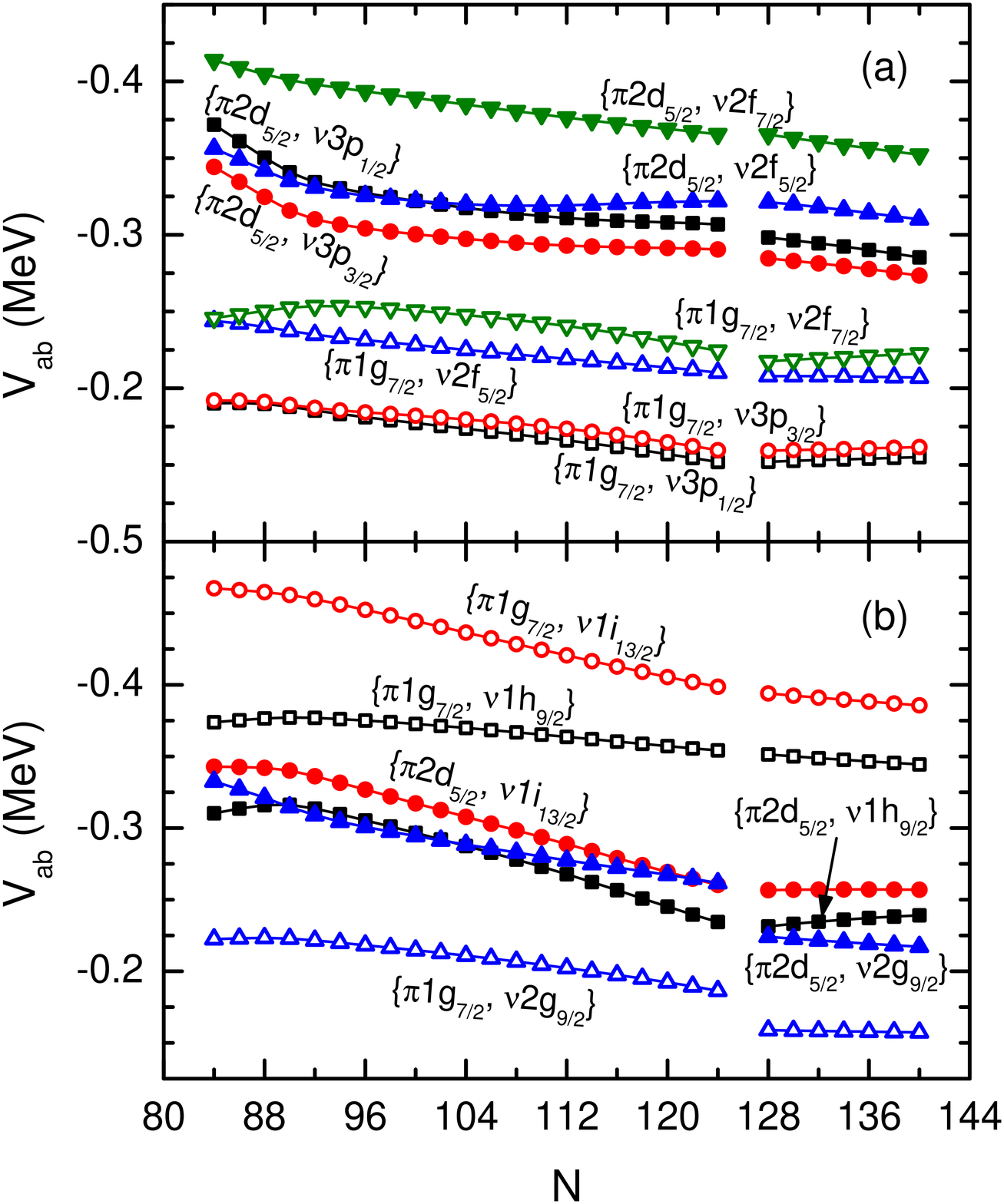}
\caption{(color online) The two-body interaction matrix elements $V_{ab}$. The notations are $a
=\pi2d_{5/2}$ (filled symbols), $\pi1g_{7/2}$ (open symbols); and $b=\nu3p_{1/2}$, $\nu3p_{3/2}$,
$\nu2f_{5/2}$, $\nu2f_{7/2}$ (a), and $b=\nu1h_{9/2}$, $\nu1i_{13/2}$, $\nu2g_{9/2}$ (b). The
results correspond to RHFB with PKA1 plus the pairing force Gogny D1S.} \label{fig:P1g2d-AT}
\end{figure}

In Fig. \ref{fig:Ceshells}b, the pseudo-spin orbital splitting $\Delta E_{\pi1\tilde f}$ given by PKA1 shows {an isospin dependence consistent with} the neutron shell evolution{, whereas in the results of PKO1 and DD-ME2} such consistency is destroyed with the violation of PSS on the pseudo-spin partner states $\pi1\tilde f$. To clarify this consistency between neutron shell evolution and proton PSS {conservation} we present in Fig. \ref{fig:P1g2d-AT} the two-body interaction matrix elements $V_{ab}$ {calculated with PKA1 and} responsible for the coupling between the proton ($a$: $\pi2d_{5/2}$ (filled symbols) and $\pi1g_{7/2}$ (open symbols)) and neutron valence orbits ($b$: $\nu2f_{7/2}$, $\nu2f_{5/2}$, $\nu3p_{3/2}$ and $\nu3p_{1/2}$ (Fig.\ref{fig:P1g2d-AT}a), and $\nu1h_{9/2}$,$\nu1i_{13/2}$ and $\nu2g_{9/2}$ (Fig.\ref{fig:P1g2d-AT}b)). It is found that the neutron orbits with nodes ($\nu2f_{7/2}$, $\nu2f_{5/2}$, $\nu3p_{3/2}$, $\nu3p_{1/2}$ and $\nu2g_{9/2}$) show a stronger coupling with the proton $\pi2d_{5/2}$ than with $\pi1g_{7/2}$ orbit, while those without node ($\nu1h_{9/2}$ and $\nu1i_{13/2}$) exhibit the opposite trend.

From $^{142}$Ce ($N=84$) to $^{148}$Ce ($N=90$) the valence neutrons are mainly filling the $\nu2f_{7/2}$ orbit, which leads to a corresponding recovery of PSS, as shown in Fig. \ref{fig:Ceshells}b. After $^{148}$Ce the orbit $\nu1h_{9/2}$, which has a stronger coupling with the $\pi1g_{7/2}$ than with the $\pi2d_{5/2}$ orbit, starts to be gradually occupied. The PSS on $\pi1\tilde f$ is still well preserved in a fairly long range from $N=90$ to $N=110$, due to the equilibrium between the states with nodes ($\nu2f_{7/2}$, $\nu2f_{5/2}$, $\nu3p_{3/2}$, $\nu3p_{1/2}$) and the one without node ($\nu1h_{9/2}$). From $N=112$ to $126$, valence neutrons are filling the $\nu1i_{13/2}$ level, as the orbits below are nearly fully occupied. An {increase} of $\Delta E_{\pi1\tilde f}$ is therefore found (see Fig. \ref{fig:Ceshells}b). When approaching the neutron drip line, the state $\nu2g_{9/2}$ starts to be occupied, and the levels below that are fully occupied. It {leads} as a result to the continuous recovery of PSS after $^{184}$Ce. In fact not only the $\nu2g_{9/2}$ state, but also the other valence orbits beyond $N=126$ {do} play a positive role in the PSS recovery. {From Fig. \ref{fig:Ceshells} and Fig. \ref{fig:P1g2d-AT} one can find that the nodal structure shows substantial effects in determining the NN interaction strength and further play an essential role in recovering the PSS. } {In fact we also found similar nodal effects as shown in Fig. \ref{fig:P1g2d-AT} from the calculations with PKO1 and DD-ME2 while the unphysically large gap between $\pi1g_{7/2}$ and $\pi2d_{5/2}$ breaks the consistency (see Fig. \ref{fig:Ceshells}) because the valence protons can only occupy the $\pi1g_{7/2}$ orbit. }


{Comparing} PKA1 to PKO1 and DD-ME2, {one finds that }the deviations on the neutron shell effects at $N=126$ in Fig. \ref{fig:Ceshells}a are mainly due to  the proton {influence}, i.e., the PSS {conservation}. In Fig. \ref{fig:P1g2d-AT}b it is seen that the protons occupying the $\pi1g_{7/2}$ level tend to enhance the neutron shell effects ($\Delta E_{N=126}$), whereas much weaker effects are contributed by particles in the $\pi2d_{5/2}$ orbit. Because of the serious violation of PSS on $\pi1\tilde f$ given by PKO1 and DD-ME2, the valence protons occupy only the $\pi1g_{7/2}$ state {and this greatly enlarges} neutron shell effects. It results in the interruption of the isotopic chain at $N=126$. In contrast the PSS is properly {conserved} by PKA1 such that the pseudo-spin partner states are simultaneously occupied and the occupations are changed consistently. It explains well the consistency between neutron shell effects and proton PSS {conservation} in Fig. \ref{fig:Ceshells}. Although some minor differences may exist on the neutron side between models, {this} would not bring any substantial changes.

 \begin{table}[htbp]
\caption{Two-body interaction matrix elements $V_{ab}$ (MeV) between proton and neutron valence
orbits in $^{198}$Ce, as well as the ratios of contributions from different coupling channels. The
four {last columns give} detailed Fock contributions, respectively
from $\rho$-vector ($\rho$-\RV), $\rho$-vector-tensor ($\rho$-\RVT), $\rho$-tensor ($\rho$-\RT) and
$\pi$-pseudo-vector ($\pi$-\RPV) couplings. The results correspond to RHFB with PKA1 plus the
pairing force Gogny D1S.} \label{tab:VAB}
\begin{tabular}{c|@{~}rrr|@{~}rrr@{~}r}\hline\hline
     &$V_{ab}$~~~&$V_{ab}^\RD$~~~&$V_{ab}^\RE$~~~&$\rho$-\RV&$\rho$-\RVT&$\rho$-\RT~~&$\pi$-\RPV\\
     \hline
 $b$ &\multicolumn{7}{c}{$a=\pi2d_{5/2}$}\\ \hline
 $\nu1i_{13/2}$&$-$0.257&63.2\%&36.8\%&9.7\%&$-$4.0\%&29.1\%&2.1\% \\ \hline
 $\nu2g_{9/2} $&$-$0.217&64.6\%&35.4\%&8.9\%&$-$3.5\%&27.0\%&3.0\% \\
 $\nu4s_{1/2} $&$-$0.097&71.8\%&28.2\%&7.1\%&$-$4.4\%&21.5\%&3.9\% \\
 $\nu3d_{5/2} $&$-$0.158&68.4\%&31.6\%&8.9\%&$-$5.4\%&25.7\%&2.5\% \\
 $\nu3d_{3/2} $&$-$0.156&67.7\%&32.3\%&7.9\%&$-$5.4\%&22.4\%&7.4\% \\
 $\nu2g_{7/2} $&$-$0.159&56.8\%&43.2\%&9.3\%&$-$8.1\%&30.8\%&11.3\%\\
 \hline\hline
 $b$ &\multicolumn{7}{c}{$a=\pi1g_{7/2}$}\\ \hline
 $\nu1i_{13/2}$&$-$0.386&63.5\%&36.5\%&8.3\%&$-$4.4\%&22.9\%&   9.8\% \\ \hline
 $\nu2g_{9/2} $&$-$0.157&64.1\%&35.9\%&8.1\%&$-$6.3\%&25.3\%&   8.8\% \\
 $\nu4s_{1/2} $&$-$0.052&68.1\%&31.9\%&8.2\%&$-$5.0\%&25.2\%&   3.5\% \\
 $\nu3d_{5/2} $&$-$0.078&67.8\%&32.2\%&7.9\%&$-$5.7\%&24.5\%&   5.6\% \\
 $\nu3d_{3/2} $&$-$0.074&70.2\%&29.8\%&8.4\%&$-$5.4\%&25.9\%&   0.9\% \\
 $\nu2g_{7/2} $&$-$0.138&69.0\%&31.0\%&9.2\%&$-$5.0\%&28.0\%&$-$1.2\% \\
 \hline\hline
\end{tabular}
\end{table}

Since the PSS {conservation} is tightly related with neutron shell effects at $N=126$, it is also essential for the stability of Cerium halo structures. Such consistency is bridged by the coupling between valence neutrons and protons, i.e., the NN interaction in $T=0$ channel. In Table \ref{tab:VAB} it is seen that Fock terms present significant contributions to the $T=0$ NN interaction, about 30\% of the total in most cases. In the $T=0$ channel, Fock terms are completely managed by isovector mesons, mainly ($\sim$2/3) by the tensor $\rho$, which can not be efficiently taken into account by the Hartree approach. From Table \ref{tab:VAB} one can find that the tensor $\rho$ plays {a} significant role not only in the conservation of PSS \cite{Long:2007} but also in bridging neutron halo structure and proton PSS recovery.

In summary, within the relativistic Hartree-Fock-Bogoliubov (RHFB) theory with density dependent meson-nucleon couplings, we studied nuclear halo phenomena occurring in Cerium isotopes and the relevant PSS {conservation}, and the role of Fock terms therein. Giant halos as well as ordinary ones are found in the drip line Ce isotopes. We also found that the stability of neutron halo structures is tightly related to the PSS {conservation} on the proton side. The Fock terms, mainly the $\rho$-tensor couplings, present substantial contributions to the NN interaction in $T=0$ channel, which accounts for the consistency between neutron halo structures and proton PSS {conservation}. In addition, the necessity of Fock terms, especially the tensor $\rho$, is well demonstrated as such effects can not be efficiently obtained with the Hartree approach.

The author {W.H.L.} would like to thank Prof. H. Sagawa for fruitful discussions. This work was supported by the Alexander von Humboldt Foundation, and Major State 973 Program 2007CB815000, as well as the National Natural Science Foundation of China under Grants No. 10435010, No. 10775004, and No. 10221003, and by the U.S.DOE grants DE-FG$\phi$2-$\phi$8ER41533 and DE-FC$\phi$2-07ER41588 (UNEDF, SciDAC-2), and by the Research Corporation.


\begin{thebibliography}{10}%
\makeatletter
\providecommand \@ifxundefined [1]{%
 \ifx #1\undefined \expandafter \@firstoftwo
 \else \expandafter \@secondoftwo
\fi
}%
\providecommand \@ifnum [1]{%
 \ifnum #1\expandafter \@firstoftwo
 \else \expandafter \@secondoftwo
\fi
}%
\providecommand \enquote [1]{``#1''}%
\providecommand \bibnamefont  [1]{#1}%
\providecommand \bibfnamefont [1]{#1}%
\providecommand \citenamefont [1]{#1}%
\providecommand\href[0]{\@sanitize\@href}%
\providecommand\@href[1]{\endgroup\@@startlink{#1}\endgroup\@@href}%
\providecommand\@@href[1]{#1\@@endlink}%
\providecommand \@sanitize [0]{\begingroup\catcode`\&12\catcode`\#12\relax}%
\@ifxundefined \pdfoutput {\@firstoftwo}{%
 \@ifnum{\z@=\pdfoutput}{\@firstoftwo}{\@secondoftwo}%
}{%
 \providecommand\@@startlink[1]{\leavevmode}%
 \providecommand\@@endlink[0]{}%
}{%
 \providecommand\@@startlink[1]{%
  \leavevmode
  \pdfstartlink
   attr{/Border[0 0 1 ]/H/I/C[0 1 1]}%
   user{/Subtype/Link/A<</Type/Action/S/URI/URI(#1)>>}%
  \relax
 }%
 \providecommand\@@endlink[0]{\pdfendlink}%
}%
\providecommand \url  [0]{\begingroup\@sanitize \@url }%
\providecommand \@url [1]{\endgroup\@href {#1}{\urlprefix}}%
\providecommand \urlprefix [0]{URL }%
\providecommand \Eprint[0]{\href }%
\@ifxundefined \urlstyle {%
  \providecommand \doi [1]{doi:\discretionary{}{}{}#1}%
}{%
  \providecommand \doi [0]{doi:\discretionary{}{}{}\begingroup
  \urlstyle{rm}\Url }%
}%
\providecommand \doibase [0]{http://dx.doi.org/}%
\providecommand \Doi[1]{\href{\doibase#1}}%
\providecommand \bibAnnote [3]{%
  \BibitemShut{#1}%
  \begin{quotation}\noindent
    \textsc{Key:}\ #2\\\textsc{Annotation:}\ #3%
  \end{quotation}%
}%
\providecommand \bibAnnoteFile [2]{%
  \IfFileExists{#2}{\bibAnnote {#1} {#2} {\input{#2}}}{}%
}%
\providecommand \typeout [0]{\immediate \write \m@ne }%
\providecommand \selectlanguage [0]{\@gobble}%
\providecommand \bibinfo [0]{\@secondoftwo}%
\providecommand \bibfield [0]{\@secondoftwo}%
\providecommand \translation [1]{[#1]}%
\providecommand \BibitemOpen[0]{}%
\providecommand \bibitemStop [0]{}%
\providecommand \bibitemNoStop [0]{.\EOS\space}%
\providecommand \EOS [0]{\spacefactor3000\relax}%
\providecommand \BibitemShut [1]{\csname bibitem#1\endcsname}%
\bibitem{Tan:1985}%
  \BibitemOpen
  \bibfield{author}{%
  \bibinfo {author} {\bibfnamefont{I.}~\bibnamefont{Tanihata}}, \bibinfo
  {author} {\bibfnamefont{H.}~\bibnamefont{Hamagaki}}, \bibinfo {author}
  {\bibfnamefont{O.}~\bibnamefont{Hashimoto}}, \bibinfo {author}
  {\bibfnamefont{Y.}~\bibnamefont{Shida}}, \bibinfo {author}
  {\bibfnamefont{N.}~\bibnamefont{Yoshikawa}}, \bibinfo {author}
  {\bibfnamefont{K.}~\bibnamefont{Sugimoto}}, \bibinfo {author}
  {\bibfnamefont{O.}~\bibnamefont{Yamakawa}}, \bibinfo {author}
  {\bibfnamefont{T.}~\bibnamefont{Kobayashi}},\ and\ \bibinfo {author}
  {\bibfnamefont{N.}~\bibnamefont{Takahashi}},\ }%
  \bibfield{journal}{%
  \bibinfo {journal} {Phys. Rev. Lett.}\ }%
  \textbf{\bibinfo {volume} {55}},\ \bibinfo {pages} {2676} (\bibinfo {year}
  {1985})%
  \bibAnnoteFile{NoStop}{Tan:1985}%
\bibitem{Howard:2000}%
  \BibitemOpen
  \bibfield{author}{%
  \bibinfo {author} {\bibfnamefont{J.~E.}\ \bibnamefont{Howard}}, \bibinfo
  {author} {\bibfnamefont{H.~R.}\ \bibnamefont{Dullin}},\ and\ \bibinfo
  {author} {\bibfnamefont{M.}~\bibnamefont{Horanyi}},\ }%
  \bibfield{journal}{%
  \bibinfo {journal} {Phys. Rev. Lett.}\ }%
  \textbf{\bibinfo {volume} {84}},\ \bibinfo {pages} {3244} (\bibinfo {year}
  {2000})%
  \bibAnnoteFile{NoStop}{Howard:2000}%
\bibitem{Mueller:2007}%
  \BibitemOpen
  \bibfield{author}{%
  \bibinfo {author} {\bibfnamefont{P.~M.}\ \bibnamefont{andI. A.~Sulai}},
  \bibinfo {author} {\bibfnamefont{A.~C.~C.}\ \bibnamefont{Villari}}, \bibinfo
  {author} {\bibfnamefont{J.~A.}\ \bibnamefont{Alcantara-Nunez}}, \bibinfo
  {author} {\bibfnamefont{R.}~\bibnamefont{Alves-Conde}}, \bibinfo {author}
  {\bibfnamefont{K.}~\bibnamefont{Bailey}}, \bibinfo {author}
  {\bibfnamefont{G.~F.}\ \bibnamefont{Drake}}, \bibinfo {author}
  {\bibfnamefont{M.}~\bibnamefont{Dubois}}, \bibinfo {author}
  {\bibfnamefont{C.}~\bibnamefont{Eleon}}, \bibinfo {author}
  {\bibfnamefont{G.}~\bibnamefont{Gaubert}}, \bibinfo {author}
  {\bibfnamefont{R.~J.}\ \bibnamefont{Holt}}, \bibinfo {author}
  {\bibfnamefont{R.~F.}\ \bibnamefont{Janssens}}, \bibinfo {author}
  {\bibfnamefont{N.}~\bibnamefont{Lecesne}}, \bibinfo {author}
  {\bibfnamefont{Z.-T.}\ \bibnamefont{Lu}}, \bibinfo {author}
  {\bibfnamefont{T.~P.}\ \bibnamefont{O'Connor}}, \bibinfo {author}
  {\bibfnamefont{M.-G.}\ \bibnamefont{Saint-Laurent}}, \bibinfo {author}
  {\bibfnamefont{J.-C.}\ \bibnamefont{Thomas}},\ and\ \bibinfo {author}
  {\bibfnamefont{L.-B.}\ \bibnamefont{Wang}},\ }%
  \bibfield{journal}{%
  \bibinfo {journal} {Phys. Rev. Lett.}\ }%
  \textbf{\bibinfo {volume} {99}},\ \bibinfo {pages} {252501} (\bibinfo {year}
  {2007})%
  \bibAnnoteFile{NoStop}{Mueller:2007}%
\bibitem{Rotival:2009a}%
  \BibitemOpen
  \bibfield{author}{%
  \bibinfo {author} {\bibfnamefont{V.}~\bibnamefont{Rotival}}\ and\ \bibinfo
  {author} {\bibfnamefont{T.}~\bibnamefont{Duguet}},\ }%
  \bibfield{journal}{%
  \bibinfo {journal} {Phys. Rev.}\ }%
  \textbf{\bibinfo {volume} {79}},\ \bibinfo {pages} {054308} (\bibinfo {year}
  {2009})%
  \bibAnnoteFile{NoStop}{Rotival:2009a}%
\bibitem{Rotival:2009b}%
  \BibitemOpen
  \bibfield{author}{%
  \bibinfo {author} {\bibfnamefont{V.}~\bibnamefont{Rotival}}, \bibinfo
  {author} {\bibfnamefont{K.}~\bibnamefont{Bennaceur}},\ and\ \bibinfo {author}
  {\bibfnamefont{T.}~\bibnamefont{Duguet}},\ }%
  \bibfield{journal}{%
  \bibinfo {journal} {Phys. Rev.}\ }%
  \textbf{\bibinfo {volume} {79}},\ \bibinfo {pages} {054309} (\bibinfo {year}
  {2009})%
  \bibAnnoteFile{NoStop}{Rotival:2009b}%
\bibitem{Dobaczewski:1994}%
  \BibitemOpen
  \bibfield{author}{%
  \bibinfo {author} {\bibfnamefont{J.}~\bibnamefont{Dobaczewski}}, \bibinfo
  {author} {\bibfnamefont{I.}~\bibnamefont{Hamamoto}}, \bibinfo {author}
  {\bibfnamefont{W.}~\bibnamefont{Nazarewicz}},\ and\ \bibinfo {author}
  {\bibfnamefont{J.~A.}\ \bibnamefont{Sheikh}},\ }%
  \bibfield{journal}{%
  \bibinfo {journal} {Phys. Rev. Lett.}\ }%
  \textbf{\bibinfo {volume} {72}},\ \bibinfo {pages} {981} (\bibinfo {year}
  {1994})%
  \bibAnnoteFile{NoStop}{Dobaczewski:1994}%
\bibitem{Chen:1995}%
  \BibitemOpen
  \bibfield{author}{%
  \bibinfo {author} {\bibfnamefont{B.}~\bibnamefont{Chen}}, \bibinfo {author}
  {\bibfnamefont{J.}~\bibnamefont{Dobaczewski}}, \bibinfo {author}
  {\bibfnamefont{K.~L.}\ \bibnamefont{Kratz}}, \bibinfo {author}
  {\bibfnamefont{K.}~\bibnamefont{Langanke}}, \bibinfo {author}
  {\bibfnamefont{B.}~\bibnamefont{Pfeiffer}}, \bibinfo {author}
  {\bibfnamefont{F.~K.}\ \bibnamefont{Thielemann}},\ and\ \bibinfo {author}
  {\bibfnamefont{P.}~\bibnamefont{Vogel}},\ }%
  \bibfield{journal}{%
  \bibinfo {journal} {Phys. Lett.}\ }%
  \textbf{\bibinfo {volume} {B 355}},\ \bibinfo {pages} {37} (\bibinfo {year}
  {1995})%
  \bibAnnoteFile{NoStop}{Chen:1995}%
\bibitem{Yukawa:1935}%
  \BibitemOpen
  \bibfield{author}{%
  \bibinfo {author} {\bibfnamefont{H.}~\bibnamefont{Yukawa}},\ }%
  \bibfield{journal}{%
  \bibinfo {journal} {Proc. Phys. Math. Soc. Japan}\ }%
  \textbf{\bibinfo {volume} {17}},\ \bibinfo {pages} {48} (\bibinfo {year}
  {1935})%
  \bibAnnoteFile{NoStop}{Yukawa:1935}%
\bibitem{Serot:1986}%
  \BibitemOpen
  \bibfield{author}{%
  \bibinfo {author} {\bibfnamefont{B.~D.}\ \bibnamefont{Serot}}\ and\ \bibinfo
  {author} {\bibfnamefont{J.~D.}\ \bibnamefont{Walecka}},\ }%
  \bibfield{journal}{%
  \bibinfo {journal} {Adv. Nucl. Phys.}\ }%
  \textbf{\bibinfo {volume} {16}},\ \bibinfo {pages} {1} (\bibinfo {year}
  {1986})%
  \bibAnnoteFile{NoStop}{Serot:1986}%
\bibitem{Otsuka:2005}%
  \BibitemOpen
  \bibfield{author}{%
  \bibinfo {author} {\bibfnamefont{T.}~\bibnamefont{Otsuka}}, \bibinfo {author}
  {\bibfnamefont{T.}~\bibnamefont{Suzuki}}, \bibinfo {author}
  {\bibfnamefont{R.}~\bibnamefont{Fujimoto}}, \bibinfo {author}
  {\bibfnamefont{H.}~\bibnamefont{Grawe}},\ and\ \bibinfo {author}
  {\bibfnamefont{Y.}~\bibnamefont{Akaishi}},\ }%
  \bibfield{journal}{%
  \bibinfo {journal} {Phys. Rev. Lett.}\ }%
  \textbf{\bibinfo {volume} {95}},\ \bibinfo {pages} {232502} (\bibinfo {year}
  {2005})%
  \bibAnnoteFile{NoStop}{Otsuka:2005}%
\bibitem{Ginocchio:2005}%
  \BibitemOpen
  \bibfield{author}{%
  \bibinfo {author} {\bibfnamefont{J.~N.}\ \bibnamefont{Ginocchio}},\ }%
  \bibfield{journal}{%
  \bibinfo {journal} {Phys. Rep.}\ }%
  \textbf{\bibinfo {volume} {414}},\ \bibinfo {pages} {165} (\bibinfo {year}
  {2005})%
  \bibAnnoteFile{NoStop}{Ginocchio:2005}%
\bibitem{Arima:1969}%
  \BibitemOpen
  \bibfield{author}{%
  \bibinfo {author} {\bibfnamefont{A.}~\bibnamefont{Arima}}, \bibinfo {author}
  {\bibfnamefont{M.}~\bibnamefont{Harvey}},\ and\ \bibinfo {author}
  {\bibfnamefont{K.}~\bibnamefont{Shimizu}},\ }%
  \bibfield{journal}{%
  \bibinfo {journal} {Phys. Lett.}\ }%
  \textbf{\bibinfo {volume} {B 30}},\ \bibinfo {pages} {517} (\bibinfo {year}
  {1969})%
  \bibAnnoteFile{NoStop}{Arima:1969}%
\bibitem{Hecht:1969}%
  \BibitemOpen
  \bibfield{author}{%
  \bibinfo {author} {\bibfnamefont{K.}~\bibnamefont{Hecht}}\ and\ \bibinfo
  {author} {\bibfnamefont{A.}~\bibnamefont{Adler}},\ }%
  \bibfield{journal}{%
  \bibinfo {journal} {Nucl. Phys.}\ }%
  \textbf{\bibinfo {volume} {A 137}},\ \bibinfo {pages} {129} (\bibinfo {year}
  {1969})%
  \bibAnnoteFile{NoStop}{Hecht:1969}%
\bibitem{Ginocchio:1997}%
  \BibitemOpen
  \bibfield{author}{%
  \bibinfo {author} {\bibfnamefont{J.~N.}\ \bibnamefont{Ginocchio}},\ }%
  \bibfield{journal}{%
  \bibinfo {journal} {Phys. Rev. Lett.}\ }%
  \textbf{\bibinfo {volume} {78}},\ \bibinfo {pages} {436} (\bibinfo {year}
  {1997})%
  \bibAnnoteFile{NoStop}{Ginocchio:1997}%
\bibitem{Meng:2006}%
  \BibitemOpen
  \bibfield{author}{%
  \bibinfo {author} {\bibfnamefont{J.}~\bibnamefont{Meng}}, \bibinfo {author}
  {\bibfnamefont{H.}~\bibnamefont{Toki}}, \bibinfo {author}
  {\bibfnamefont{S.~G.}\ \bibnamefont{Zhou}}, \bibinfo {author}
  {\bibfnamefont{S.~Q.}\ \bibnamefont{Zhang}}, \bibinfo {author}
  {\bibfnamefont{W.~H.}\ \bibnamefont{Long}},\ and\ \bibinfo {author}
  {\bibfnamefont{L.~S.}\ \bibnamefont{Geng}},\ }%
  \bibfield{journal}{%
  \bibinfo {journal} {Prog. Part. Nucl. Phys.}\ }%
  \textbf{\bibinfo {volume} {57}},\ \bibinfo {pages} {470} (\bibinfo {year}
  {2006})%
  \bibAnnoteFile{NoStop}{Meng:2006}%
\bibitem{Meng:1998a}%
  \BibitemOpen
  \bibfield{author}{%
  \bibinfo {author} {\bibfnamefont{J.}~\bibnamefont{Meng}},\ }%
  \bibfield{journal}{%
  \bibinfo {journal} {Nucl. Phys.}\ }%
  \textbf{\bibinfo {volume} {A 635}},\ \bibinfo {pages} {3} (\bibinfo {year}
  {1998})%
  \bibAnnoteFile{NoStop}{Meng:1998a}%
\bibitem{Vretenar:2005}%
  \BibitemOpen
  \bibfield{author}{%
  \bibinfo {author} {\bibfnamefont{D.}~\bibnamefont{Vretenar}}, \bibinfo
  {author} {\bibfnamefont{A.~V.}\ \bibnamefont{Afanasjev}}, \bibinfo {author}
  {\bibfnamefont{G.~A.}\ \bibnamefont{Lalazissis}},\ and\ \bibinfo {author}
  {\bibfnamefont{P.}~\bibnamefont{Ring}},\ }%
  \bibfield{journal}{%
  \bibinfo {journal} {Phys. Rep.}\ }%
  \textbf{\bibinfo {volume} {409}},\ \bibinfo {pages} {101} (\bibinfo {year}
  {2005})%
  \bibAnnoteFile{NoStop}{Vretenar:2005}%
\bibitem{Meng:1996}%
  \BibitemOpen
  \bibfield{author}{%
  \bibinfo {author} {\bibfnamefont{J.}~\bibnamefont{Meng}}\ and\ \bibinfo
  {author} {\bibfnamefont{P.}~\bibnamefont{Ring}},\ }%
  \bibfield{journal}{%
  \bibinfo {journal} {Phys. Rev. Lett.}\ }%
  \textbf{\bibinfo {volume} {77}},\ \bibinfo {pages} {3963} (\bibinfo {year}
  {1996})%
  \bibAnnoteFile{NoStop}{Meng:1996}%
\bibitem{Poschl:1997}%
  \BibitemOpen
  \bibfield{author}{%
  \bibinfo {author} {\bibfnamefont{W.}~\bibnamefont{P\"oschl}}, \bibinfo
  {author} {\bibfnamefont{D.}~\bibnamefont{Vretenar}}, \bibinfo {author}
  {\bibfnamefont{G.~A.}\ \bibnamefont{Lalazissis}},\ and\ \bibinfo {author}
  {\bibfnamefont{P.}~\bibnamefont{Ring}},\ }%
  \bibfield{journal}{%
  \bibinfo {journal} {Phys. Rev. Lett.}\ }%
  \textbf{\bibinfo {volume} {79}},\ \bibinfo {pages} {3841} (\bibinfo {year}
  {1997})%
  \bibAnnoteFile{NoStop}{Poschl:1997}%
\bibitem{Lalazissis:1998NPA}%
  \BibitemOpen
  \bibfield{author}{%
  \bibinfo {author} {\bibfnamefont{G.~A.}\ \bibnamefont{Lalazissis}}, \bibinfo
  {author} {\bibfnamefont{D.}~\bibnamefont{Vretenar}}, \bibinfo {author}
  {\bibfnamefont{W.}~\bibnamefont{P\"oschl}},\ and\ \bibinfo {author}
  {\bibfnamefont{P.}~\bibnamefont{Ring}},\ }%
  \bibfield{journal}{%
  \bibinfo {journal} {Nucl. Phys.}\ }%
  \textbf{\bibinfo {volume} {A 632}},\ \bibinfo {pages} {363} (\bibinfo {year}
  {1998})%
  \bibAnnoteFile{NoStop}{Lalazissis:1998NPA}%
\bibitem{Meng:1998PLB}%
  \BibitemOpen
  \bibfield{author}{%
  \bibinfo {author} {\bibfnamefont{J.}~\bibnamefont{Meng}}, \bibinfo {author}
  {\bibfnamefont{I.}~\bibnamefont{Tanihata}},\ and\ \bibinfo {author}
  {\bibfnamefont{S.}~\bibnamefont{Yamaji}},\ }%
  \bibfield{journal}{%
  \bibinfo {journal} {Phys. Lett.}\ }%
  \textbf{\bibinfo {volume} {B 419}},\ \bibinfo {pages} {1} (\bibinfo {year}
  {1998})%
  \bibAnnoteFile{NoStop}{Meng:1998PLB}%
\bibitem{Meng:2002PRC}%
  \BibitemOpen
  \bibfield{author}{%
  \bibinfo {author} {\bibfnamefont{J.}~\bibnamefont{Meng}}, \bibinfo {author}
  {\bibfnamefont{H.}~\bibnamefont{Toki}}, \bibinfo {author}
  {\bibfnamefont{J.~Y.}\ \bibnamefont{Zeng}}, \bibinfo {author}
  {\bibfnamefont{S.~Q.}\ \bibnamefont{Zhang}},\ and\ \bibinfo {author}
  {\bibfnamefont{S.~G.}\ \bibnamefont{Zhou}},\ }%
  \bibfield{journal}{%
  \bibinfo {journal} {Phys. Rev.}\ }%
  \textbf{\bibinfo {volume} {C 65}},\ \bibinfo {pages} {041302} (\bibinfo
  {year} {2002})%
  \bibAnnoteFile{NoStop}{Meng:2002PRC}%
\bibitem{Meng:1998PRL}%
  \BibitemOpen
  \bibfield{author}{%
  \bibinfo {author} {\bibfnamefont{J.}~\bibnamefont{Meng}}\ and\ \bibinfo
  {author} {\bibfnamefont{P.}~\bibnamefont{Ring}},\ }%
  \bibfield{journal}{%
  \bibinfo {journal} {Phys. Rev. Lett.}\ }%
  \textbf{\bibinfo {volume} {80}},\ \bibinfo {pages} {460} (\bibinfo {year}
  {1998})%
  \bibAnnoteFile{NoStop}{Meng:1998PRL}%
\bibitem{Grasso:2006PRC}%
  \BibitemOpen
  \bibfield{author}{%
  \bibinfo {author} {\bibfnamefont{M.}~\bibnamefont{Grasso}}, \bibinfo {author}
  {\bibfnamefont{S.}~\bibnamefont{Yoshida}}, \bibinfo {author}
  {\bibfnamefont{N.}~\bibnamefont{Sandulescu}},\ and\ \bibinfo {author}
  {\bibfnamefont{N.~V.}\ \bibnamefont{Giai}},\ }%
  \bibfield{journal}{%
  \bibinfo {journal} {Phys. Rev.}\ }%
  \textbf{\bibinfo {volume} {C 74}},\ \bibinfo {pages} {064317} (\bibinfo
  {year} {2006})%
  \bibAnnoteFile{NoStop}{Grasso:2006PRC}%
\bibitem{Long:2006}%
  \BibitemOpen
  \bibfield{author}{%
  \bibinfo {author} {\bibfnamefont{W.~H.}\ \bibnamefont{Long}}, \bibinfo
  {author} {\bibfnamefont{N.~V.}\ \bibnamefont{Giai}},\ and\ \bibinfo {author}
  {\bibfnamefont{J.}~\bibnamefont{Meng}},\ }%
  \bibfield{journal}{%
  \bibinfo {journal} {Phys. Lett.}\ }%
  \textbf{\bibinfo {volume} {B 640}},\ \bibinfo {pages} {150} (\bibinfo {year}
  {2006})%
  \bibAnnoteFile{NoStop}{Long:2006}%
\bibitem{Long:2007}%
  \BibitemOpen
  \bibfield{author}{%
  \bibinfo {author} {\bibfnamefont{W.~H.}\ \bibnamefont{Long}}, \bibinfo
  {author} {\bibfnamefont{H.}~\bibnamefont{Sagawa}}, \bibinfo {author}
  {\bibfnamefont{N.~V.}\ \bibnamefont{Giai}},\ and\ \bibinfo {author}
  {\bibfnamefont{J.}~\bibnamefont{Meng}},\ }%
  \bibfield{journal}{%
  \bibinfo {journal} {Phys. Rev.}\ }%
  \textbf{\bibinfo {volume} {C 76}},\ \bibinfo {pages} {034314} (\bibinfo
  {year} {2007})%
  \bibAnnoteFile{NoStop}{Long:2007}%
\bibitem{Long:2008}%
  \BibitemOpen
  \bibfield{author}{%
  \bibinfo {author} {\bibfnamefont{W.~H.}\ \bibnamefont{Long}}, \bibinfo
  {author} {\bibfnamefont{H.}~\bibnamefont{Sagawa}}, \bibinfo {author}
  {\bibfnamefont{J.}~\bibnamefont{Meng}},\ and\ \bibinfo {author}
  {\bibfnamefont{N.~V.}\ \bibnamefont{Giai}},\ }%
  \bibfield{journal}{%
  \bibinfo {journal} {Europhysics Letters}\ }%
  \textbf{\bibinfo {volume} {82}},\ \bibinfo {pages} {12001} (\bibinfo {year}
  {2008})%
  \bibAnnoteFile{NoStop}{Long:2008}%
\bibitem{Long:2006PS}%
  \BibitemOpen
  \bibfield{author}{%
  \bibinfo {author} {\bibfnamefont{W.~H.}\ \bibnamefont{Long}}, \bibinfo
  {author} {\bibfnamefont{H.}~\bibnamefont{Sagawa}}, \bibinfo {author}
  {\bibfnamefont{J.}~\bibnamefont{Meng}},\ and\ \bibinfo {author}
  {\bibfnamefont{N.~V.}\ \bibnamefont{Giai}},\ }%
  \bibfield{journal}{%
  \bibinfo {journal} {Phys. Lett.}\ }%
  \textbf{\bibinfo {volume} {B 639}},\ \bibinfo {pages} {242} (\bibinfo {year}
  {2006})%
  \bibAnnoteFile{NoStop}{Long:2006PS}%
\bibitem{Long:2009b}%
  \BibitemOpen
  \bibfield{author}{%
  \bibinfo {author} {\bibfnamefont{W.~H.}\ \bibnamefont{Long}}, \bibinfo
  {author} {\bibfnamefont{P.}~\bibnamefont{Ring}}, \bibinfo {author}
  {\bibfnamefont{N.~V.}\ \bibnamefont{Giai}},\ and\ \bibinfo {author}
  {\bibfnamefont{J.}~\bibnamefont{Meng}}}%
   (\bibinfo {year} {2009}),\
  \Eprint{http://arxiv.org/abs/nucl-th/0812.1103}{nucl-th/0812.1103}%
  \bibAnnoteFile{NoStop}{Long:2009b}%
\bibitem{Nagai:1981}%
  \BibitemOpen
  \bibfield{author}{%
  \bibinfo {author} {\bibfnamefont{Y.}~\bibnamefont{Nagai}}, \bibinfo {author}
  {\bibfnamefont{J.}~\bibnamefont{Styczen}}, \bibinfo {author}
  {\bibfnamefont{M.}~\bibnamefont{Piiparinen}},\ and\ \bibinfo {author}
  {\bibfnamefont{P.}~\bibnamefont{Kleinheinz}},\ }%
  \bibfield{journal}{%
  \bibinfo {journal} {Phys. Rev. Lett.}\ }%
  \textbf{\bibinfo {volume} {47}},\ \bibinfo {pages} {1259} (\bibinfo {year}
  {1981})%
  \bibAnnoteFile{NoStop}{Nagai:1981}%
\bibitem{Long:2009}%
  \BibitemOpen
  \bibfield{author}{%
  \bibinfo {author} {\bibfnamefont{W.~H.}\ \bibnamefont{Long}}, \bibinfo
  {author} {\bibfnamefont{T.}~\bibnamefont{Nakatsukasa}}, \bibinfo {author}
  {\bibfnamefont{H.}~\bibnamefont{Sagawa}}, \bibinfo {author}
  {\bibfnamefont{J.}~\bibnamefont{Meng}}, \bibinfo {author}
  {\bibfnamefont{H.}~\bibnamefont{Nakada}},\ and\ \bibinfo {author}
  {\bibfnamefont{Y.}~\bibnamefont{Zhang}},\ }%
  \bibfield{journal}{%
  \bibinfo {journal} {Phys. Lett.}\ }%
  \textbf{\bibinfo {volume} {B 680}},\ \bibinfo {pages} {428} (\bibinfo {year}
  {2009})%
  \bibAnnoteFile{NoStop}{Long:2009}%
\bibitem{Berger84}%
  \BibitemOpen
  \bibfield{author}{%
  \bibinfo {author} {\bibfnamefont{J.~F.}\ \bibnamefont{Berger}}, \bibinfo
  {author} {\bibfnamefont{M.}~\bibnamefont{Girod}},\ and\ \bibinfo {author}
  {\bibfnamefont{D.}~\bibnamefont{Gogny}},\ }%
  \bibfield{journal}{%
  \bibinfo {journal} {Nucl. Phys.}\ }%
  \textbf{\bibinfo {volume} {A 428}},\ \bibinfo {pages} {23} (\bibinfo {year}
  {1984})%
  \bibAnnoteFile{NoStop}{Berger84}%
\bibitem{Lalazissis:2005}%
  \BibitemOpen
  \bibfield{author}{%
  \bibinfo {author} {\bibfnamefont{G.~A.}\ \bibnamefont{Lalazissis}}, \bibinfo
  {author} {\bibfnamefont{T.}~\bibnamefont{Nik$\check{\rm s}$i$\acute{\rm
  c}$}}, \bibinfo {author} {\bibfnamefont{D.}~\bibnamefont{Vretenar}},\ and\
  \bibinfo {author} {\bibfnamefont{P.}~\bibnamefont{Ring}},\ }%
  \bibfield{journal}{%
  \bibinfo {journal} {Phys. Rev.}\ }%
  \textbf{\bibinfo {volume} {C 71}},\ \bibinfo {pages} {024312} (\bibinfo
  {year} {2005})%
  \bibAnnoteFile{NoStop}{Lalazissis:2005}%
\end{thebibliography}

%

\end{document}